# Mechanism of multisoliton formation and soliton energy quantization in passively mode-locked fiber lasers

D. Y. Tang, L. M. Zhao, B. Zhao and A. Q. Liu

School of Electrical and Electronic Engineering, Nanyang Technological University, Singapore

## Abstract

We report results of numerical simulations on the multiple soliton generation and soliton energy quantization in a soliton fiber ring laser passively mode-locked by using the nonlinear polarization rotation technique. We found numerically that the formation of multiple solitons in the laser is caused by a peak power limiting effect of the laser cavity. It is also the same effect that suppresses the soliton pulse collapse, an intrinsic feature of solitons propagating in the gain media, and makes the solitons stable in the laser. Furthermore, we show that the soliton energy quantization observed in the lasers is a natural consequence of the gain competition between the multiple solitons. Enlightened by the numerical result we speculate that the multi-soliton formation and soliton energy quantization observed in other types of soliton fiber lasers could have similar mechanism.

## 1. Introduction

Passively mode locked fiber lasers as a simple and economic ultrashort pulse source have been extensively investigated in the past decade [1-9]. By implementing the soliton pulse shaping technique in the lasers it was demonstrated that optical pulses in the sub-picosecond range could be routinely generated. Various passive mode-locking techniques, such as the nonlinear loop mirror method [3, 4], the nonlinear polarization rotation (NPR) technique [5-7] and the semiconductor saturable absorber method [8, 9], have been used to mode-lock the lasers. Independent of the concrete mode-locking techniques it was found that the soliton operation of all the lasers exhibited a common feature, namely under strong pumping strength multiple soliton pulses are always generated in the laser cavity, and in the steady state all the solitons have exactly the same pulse properties: the same pulse energy and pulse width when they are far apart. The latter property of the solitons was also called as the "soliton energy quantization effect" [10]. The multiple soliton generation and the soliton energy quantization effect limit the generation of optical pulses with larger pulse energy and narrower pulse width in the lasers. Therefore, in order to further improve the performance of the lasers it is essential to have a clear understanding on the physical mechanism responsible for these effects. It was conjectured that the soliton energy quantization could be an intrinsic property of the laser solitons, as solitons formed in a laser are intrinsically dissipative solitons, where the requirement of soliton internal energy balance ultimately determines the energy of a soliton [11]. However, this argument can not explain the formation of multiple solitons in a laser cavity. Actually multiple pulse generation have also been observed in other type of soliton lasers, e.g. M. J. Lederer *et. al.* reported the multipulse operation of a Ti:sapphire laser mode-locked by an ion-implanted semiconductor saturable absorber mirror [12], C. Spielmann *et. al.* reported the breakup of single pulses into multiple pulses in a Kerr lens mode-locked Ti:sapphire laser [13]. Theoretically, Kärtner *et. al.* have proposed a mechanism of pulse splitting for the multiple pulse generation in soliton lasers [14]. It was shown that when a pulse in a laser becomes so narrow that due to the effective gain bandwidth limit, the gain could no longer amplify the pulse but impose an extra loss on it, the pulse would split into two pulses with broader pulse width. Based on a similar mechanism and in the framework of a generalized complex Ginzburg-Landau equation that explicitly takes into account the effect of a bandpass filter in the cavity, M.J. Lederer theoretically explained the multiple pulse operation of their laser [12]. However, we point out that this process of multipulse generation can be easily identified experimentally. In the case that no bandpass filter is in the cavity, a pulse splits into two pulses only when its pulse width has become so narrow that it is limited by the gain bandwidth. While in the case of fiber lasers no significant soliton pulse narrowing was ever observed before a new soliton pulse was generated, which obviously demonstrated that the multiple pulse generation in the soliton fiber lasers must have a different mechanism. G. P. Agrawal has also numerically shown multiple pulse formation when a pulse propagates in a strongly pumped gain medium [15]. Nevertheless, it can be shown that the multiple pulse formation has in fact the same mechanism as that described by Kärtner *et. al*. Recently, P. Grelu et. al. have numerically simulated the multiple pulse operation of the fiber soliton lasers [16,17]. By using a propagation model and also taking into account the laser cavity effect, they could quite well reproduce the multipulse states of the experimental observations. However, no analysis on the physical mechanism of the multipulse formation was given. In addition, in their simulations the multisoliton formation is only obtained for limited sets of parameters, which is not in agreement with the experimental observations. Very recently, A. Komarov et. al. have theoretically studied the multiple soliton operation and pump hysteresis of the soliton fiber lasers mode-locked by using the NPR technique [18]. In their model they have explicitly taken into account the nonlinear cavity effect so they can successfully explain the multisoliton formation and pump hysteresis based on the nonlinear cavity feedback. However, as they ignored the linear birefringence of the fiber and the associated linear cavity effects, their model could still not accurately describe the real laser systems, e.g., in their model in order to obtain the multiple pulse operation, they have to add phenomenologically a frequency selective loss term. Physically, adding the term is like to add a bandpass filter in the laser cavity.

In this paper we present results of numerical simulations on the soliton formation and soliton energy quatization in a fiber ring laser passively mode-locked by using the NPR technique. First we show that soliton formation is actually a natural consequence of a mode-locked pulse under strong pumping if a laser is operating in the anomalous total cavity dispersion regime. Especially we will show how the parameters of a laser soliton, such as the peak power and pulse width, vary with the laser operation conditions. Based on our numerical simulations we further show that, for the first time to our knowledge, the multiple soliton formation in the laser is caused by a peak power limiting effect of the laser cavity. It is also the effect of the cavity that suppresses the soliton collapse and makes the solitons stable in the laser even when the laser gain is very strong. Furthermore, we demonstrate numerically that the soliton energy quantization of the laser is a nature consequence of the gain competition between the solitons in the cavity.

## 2. Experimental observations

For the purpose of comparison and a better understanding of our numerical simulations, we present here again some of the typical experimental results on the multiple soliton operation and soliton energy quantization of the soliton fiber lasers. We note that although the results presented here were obtained from a particular soliton fiber ring laser as described below, similar features were also observed in other lasers [3,4,5,8], which are in fact independent of the concrete laser systems. A schematic of the fiber soliton laser we used in our experiments is shown in Fig. 1. It contains 1-meter-long dispersion shifted fiber with group velocity dispersion (GVD) of about −2 ps/nm km, 4-meter-long erbium doped fiber (EDF) with GVD of about −10 ps/nm km and 1-meter-long standard single mode fiber (SMF) with GVD of about −18 ps/nm km. Two polarization controllers (PCs), one consisting of two quarter-wave plates and the other one two quarter-wave plates and one half-wave plate, were used to control the polarization of the light in the cavity. A polarization-dependent isolator was used to enforce the unidirectional operation of the laser and also determine the polarization of the light at the position. A 10% output-coupler was used to outlet the light. The soliton pulse width of the laser was measured with a commercial autocorrelator, and the average soliton output power was measured with a power meter. The soliton pulse evolution inside the laser cavity was monitored with a high-speed detector and a sampling oscilloscope.

The soliton operation of the laser was extensively investigated previously [19-21], various features such as the pump power hysteresis, multiple soliton generation and various modes of multiple soliton operation, bound states of solitons were observed. Worth of mentioning here again is the pump hysteresis effect of the soliton operation. It was found experimentally that the laser always started mode locking at a high pump power level, and immediately after the mode-locking multiple solitons were formed in the cavity. After the soliton operation was obtained, the laser pump power could then be reduced to a very low level while the laser still maintained the soliton operation. This phenomenon of the laser soliton operation was known as the pump power hysteresis [22]. It was later turned out that the pump power hysteresis effect is related to the multiple soliton operation of the laser. Once multiple solitons are generated in the cavity, decreasing the pump power the number of solitons is reduced. However, as far as one soliton is remained in the cavity, the soliton operation state (and therefore the mode-locking of the laser) is maintained. Not only the soliton operation of the laser exhibited pump power hysteresis, but also the generation and annihilation of each individual soliton in the laser exhibited pump power hysteresis [19]. Experimentally it was observed that if there were already solitons in the cavity, carefully increasing the pump power, new solitons could be generated one by one in the cavity. As in this case the laser is already mode-locked, the generation of a new soliton only requires a small increase of the pump power.

An important characteristic of the multiple soliton operation of the laser is that, as far as the solitons are far apart in the cavity, they all have exactly the same soliton parameters: the same pulse width, pulse energy and peak power. To demonstrate the property we have shown in Fig. 2 the oscilloscope trace of a typical experimentally measured multiple soliton operation state of our laser. The cavity round trip time of the laser is about 26ns. There are 6 solitons coexisting in the cavity. It can be clearly seen that each soliton has exactly the same pulse height in the oscilloscope trace. Although with the electronic detection system the detailed pulse profile of the solitons cannot be resolved, nevertheless, the measured pulse height in the oscilloscope trace is directly related to the energy of each individual solitons. Based on the measured autocorrelation traces and optical spectra it was further identified that all the solitons indeed have exactly the same soliton parameters.

## 3. Theoretical modeling

To find out the physical mechanism of the multiple soliton formation in our laser, we have experimentally carefully investigated its soliton operation and compared with those calculated from the conventional models of the fiber soliton lasers. Traditionally, the soliton operation of a laser was modeled by the Ginzburg-Landau equation [23] or the Master equation [24], which also takes into account the gain, loss and saturable absorber effects of a laser. However, a drawback of the models is that the laser cavity effect on the soliton was either ignored or not appropriately considered. Based on results of our experimental studies, we found that the cavity properties affect significantly the features of the solitons as in the case of soliton lasers the solitons circulate inside a cavity. Therefore, we believe that in order to accurately model the soliton operation of a laser, the detailed cavity property must also be included in the model. To this end we have extended the conventional Ginzburg-Landau equation model through incorporating the cavity features. In previous papers we have reported results of using the model in simulating the experimentally observed soliton sideband asymmetry [25], sub-sideband generation [26], twin-pulse soliton [27] and soliton pulse train non-uniformity [28]. We found that with the new model we could well reproduce all the experimentally observed phenomena of our laser.

The basic idea of our model, which is fundamentally different to the conventional models, is that we didn't make the small pulse variation approximation. Instead we follow the circulation of the optical pulses in the laser cavity and consider every action of the cavity components on the pulses. Concretely, we describe the light propagation in the optical fibers by the nonlinear Schrödinger equation, or the coupled nonlinear Schrödinger equation if the fiber is weakly birefringent. For the erbium-doped fiber, we also incorporate the gain effect such as the light amplification and gain bandwidth limitation in the equation. Whenever the pulse encounters a discrete cavity component, e.g. the output coupler, polarizer, we then account the effect of the cavity component by multiply its transfer matrix to the light field. As the model itself is very complicated, we have to numerically solve it and find out the eigenstate of the laser under a certain operation condition. In our numerical simulations we always start the calculation with an arbitrary light field. After one round trip circulation in the cavity, we then use the calculated result as the input of the next round of calculation until a steady state is obtained. We found that the simulations will always approach to a stable solution, which corresponds to a stable laser state under a certain operation condition.

To illustrate our technique, we present here the detailed procedure in simulating the soliton operation of the laser described above. To describe the light propagation in the weakly birefringent fibers, we used the coupled complex nonlinear Schrödinger equations of the form:

$$\begin{cases} \dfrac{\partial u}{\partial z} = -i\beta u + \delta \dfrac{\partial u}{\partial t} - \dfrac{ik''}{2}\dfrac{\partial^2 u}{\partial t^2} + \dfrac{ik'''}{6}\dfrac{\partial^3 u}{\partial t^3} + i\gamma(|u|^2 + \dfrac{2}{3}|v|^2)u + \dfrac{i\gamma}{3}v^2 u^* + \dfrac{g}{2}u + \dfrac{g}{2\Omega_g}\dfrac{\partial^2 u}{\partial t^2} \\ \dfrac{\partial v}{\partial z} = i\beta v - \delta \dfrac{\partial v}{\partial t} - \dfrac{ik''}{2}\dfrac{\partial^2 v}{\partial t^2} + \dfrac{ik'''}{6}\dfrac{\partial^3 v}{\partial t^3} + i\gamma(|v|^2 + \dfrac{2}{3}|u|^2)v + \dfrac{i\gamma}{3}u^2 v^* + \dfrac{g}{2}v + \dfrac{g}{2\Omega_g}\dfrac{\partial^2 v}{\partial t^2} \end{cases} \quad (1)$$

where $u$ and $v$ are the normalized envelopes of the optical pulses along the two orthogonal polarization axes of the fiber. $2\beta = 2\pi\Delta n/\lambda$ is the wave-number difference between the two modes. $2\delta = 2\beta\lambda/2\pi c$ is the inverse group velocity difference. $k''$ is the second order dispersion coefficient, $k'''$ is the third order dispersion coefficient and $\gamma$ represents the nonlinearity of the fiber. $g$ is the saturable gain of the fiber and $\Omega_g$ is the bandwidth of the laser gain. For undoped fibers $g = 0$, for the erbium-doped fiber, we further considered the gain saturation as

$$g = G\exp[-\dfrac{\int(|u|^2 + |v|^2)dt}{P_{sat}}] \quad (2)$$

where G is the small signal gain coefficient and $P_{sat}$ is the normalized saturation energy.

To possibly close to the experimental conditions of our laser, we have used the following fiber parameters for our simulations: $\gamma$=3 W$^{-1}$km$^{-1}$, $k'''$ =0.1 ps$^2$/nm km, $\Omega_g$ = **20 nm**, gain saturation energy P$_{sat}$ =1000, cavity length L=6m and the beat length of the fiber birefringence L$_b$=L/4. To simulate the cavity effect, we let the light circulate in the cavity. Starting from the intra-cavity polarizer, which has an orientation of θ = 0.125π to the fiber's fast axis, the light then propagates in the various fibers, first through the 1m dispersion shifted fiber (DSF), which has a GVD coefficient of $k''$ = -2ps/nm/km, then the 4m EDF whose GVD coefficient $k''$ = -10ps/nm/km, and finally the 1m standard single mode fiber whose GVD coefficient $k''$ = -18 ps/nm/km. Subsequently the light passes through the waveplates, which cause a fixed polarization rotation of the light. Note that changing the relative orientations of the waveplates is physically equivalent to add a variable linear cavity phase delay bias to the cavity. Certainly the principal polarization axes of the waveplates are not aligned with those of the fibers, and in general the different fibers used in the laser cavity could also have different principal polarization axes. However, for the simplicity of numerical calculations, we have treated them all having the same principal polarization axes, and considered the effect caused by the principal polarization axis change by assuming that the polarizer has virtually a different orientation to the fast axis of the fiber when it acts as an analyzer. In our simulations the orientation angle of the analyzer to the fibers fast axis is set as φ =π/2+θ. It is to point out that in the real laser system the analyzer is also the same polarizer (PI). Therefore, the light after the analyzer is also the light after the polarizer. We then used the light as the input for the next round calculation, and the procedure repeats until a steady state is achieved.

## 4. Simulation results

The coupled complex nonlinear Schrödinger equations (1) were numerically solved by using the split-step Fourier method [29]. We found that by appropriately setting the linear cavity phase delay bias of the cavity, so that an artificial saturable absorber effect can be generated in the laser, self-started mode locking can always be generated in our simulations through simply increasing the small signal gain coefficient, which corresponds to increasing the pump power in the experiments. Exactly like the experimental observations, multiple soliton pulses are formed in the simulation window immediately after the mode locking. In the steady state and when the solitons are far separated, all the solitons obtained have exactly the same pulse parameters such as the peak power and pulse width. Fig. 3 shows for example a numerically calculated multiple soliton operation of the laser. Like the experimental observations, the soliton operation of the laser and the generation and annihilation of each individual soliton in the cavity exhibit pump hystersis. Decreasing numerically the pump power, the soliton number in the simulation window reduces one by one, while carefully increasing the pump strength, with at least one soliton already existing in the cavity, solitons can also be generated one by one as shown in fig. 4. All these numerically calculated results are in excellent agreement with the experimental observations.

In a practical laser due to the existence of laser output, fiber splices etc, the linear cavity losses are unavoidable. However, in the numerical simulations we could artificially reduce the linear cavity losses and even make it to zero. We found numerically that the weaker the linear cavity loss, the smaller is the pump hysteresis of the soliton operation. With a very weak linear cavity loss we found numerically that a single soliton pulse could even be directly formed from a mode locked pulse through increasing the pump strength. This numerical result clearly shows that the large pump hysteresis of the soliton operation of the laser is caused by the existence of large linear cavity loss of a practical laser. A large linear cavity loss makes the mode locking threshold of a laser very high, which under the existence of cavity saturable absorber effect, causes that the effective gain of the laser after mode-locking is very large. As will be shown below, when the peak power of a pulse is clamped, this large effective laser gain will then results in the formation of multiple solitons immediately after the mode-locking of the laser. We have mentioned in the introduction part the theoretical work of A. Komarov et. al. on the multistability and hysteresis phenomena in passively mode-locked fiber lasers. In the framework of their model they have explained these phenomena as caused by the competition between the positive nonlinear feedback and the negative phase modulation effect [18]. It is to note that in their model in order to obtain the multiple soliton operation, the cavity loss term caused by the frequency selective filter has to be added in, which from another aspect confirms our numerical result shown above.

By making the linear cavity loss small, we have numerically investigated the process on how a soliton is formed in the laser cavity. Fig. 5 shows the results of numerical simulations. In obtaining the result the linear cavity phase delay bias is set to $\delta\Phi_l=1.2\pi$. When G is less than 251, there is no mode locking. In the experiment this corresponds to the case that the laser is operating below the mode-locking threshold. When G is equal to 252, a mode-locked pulse emerges in the cavity. The mode-locked pulse has weak pulse intensity and broad pulse width. Due to the action of the mode locker, which in the laser is the artificial saturable absorber, the mode-locked pulse circulates stably in the cavity, just like any mode-locked pulse in other lasers. Although such a mode-locked pulse has stable pulse profile during circulation in the cavity, we emphasize that it is not a soliton but a linear pulse. The linear nature of the pulse is also reflected by that its optical spectrum has no sidebands. When G is further increased, the peak power of the pulse quickly increases. Associated with the pulse intensity increase the nonlinear optical Kerr effect of the fiber also becomes strong and eventually starts to play a role. An effect of the pulse self-phase-modulation (SPM) is to generate a positive frequency chirp, which in the anomalous cavity dispersion regime counterbalances the negative frequency chirp caused by the cavity dispersion effect and compresses the pulse width. When the pulse peak power has become so strong that the nonlinear SPM effect alone can balance the pulse broadening caused by the cavity dispersion effect, even without the existence of the mode-locker, a pulse can propagate stably in the dispersive laser cavity. In this case a mode-locked pulse then becomes a soliton. In the case of our simulation, this corresponds to the state of G=253. A soliton in the laser is also characterized by the appearance of the sidebands in the optical spectrum as shown in Fig. 5.

Once the laser gain is fixed, a soliton with fixed peak power and pulse width will be formed, which are independent of the initial conditions. The states shown in Fig. 5 are stable and unique. This result confirms the auto-soliton property of the laser solitons [30]. However, if the pump power is continuously increased, solitons with even higher peak power and narrower pulse width will be generated. Associated with the soliton pulse width narrowing, the spectrum of the soliton broadens, and consequently more sidebands become visible. However, the positions of the sidebands are almost fixed. The physical mechanism of sideband generation of laser solitons was extensively investigated previously and is well understood now [31]. It is widely believed that the sideband generation is a fundamental limitation to the soliton pulse narrowing in a laser [32]. However, our numerical simulations clearly show that the sideband generation is just an adaptive effect, whose existence does not limit the soliton pulse narrowing. As far as the pump power could balance the loss caused by the sidebands, soliton pulse width can still be narrowed. Based on our numerical simulation and if there is no other limitation as will be described below in the paper, the narrowest soliton pulse that can be formed in a laser should be ultimately only determined by the laser cavity dispersion property, including the net dispersion of all the cavity components and the dispersion of the gain medium.

With already one soliton in the simulation window, we then further increased the pump strength. Depending on the selection of the linear cavity phase delay bias, we found that the mechanism of the new soliton generation and the features of the multiple soliton operation in the laser are different. With the laser parameters as described above, we found that when the linear cavity phase delay bias is set small, say at about $\delta\Phi_l=1.2\pi$, further increasing the pump power, initially the soliton pulse peak power will be increased and its pulse width narrowed as expected. However, to a certain fixed value these will stop, instead the background of the simulation window becomes unstable and weak background pulses become visible as shown in figure 6b. Further slightly increasing the pump power, a new soliton is quickly formed in the cavity through the soliton shaping of one of the weak background pulses. As the weak background pulses are always initiated from the dispersive waves of the solitons, we have called the new soliton generation "soliton shaping of dispersive waves" [33]. In the steady state both solitons have exactly the same pulse width and peak power as shown in figure 6c. When the pump power is further increased, new solitons are generated one by one in the simulation window in exactly the same way and eventually a multiple soliton state as shown in Fig. 3 is obtained. This numerically simulated result is well in agreement with the experimental observations [19]. Because of the new soliton generation, the solitons formed in the laser cannot have large pulse energy and high peak power through simply increasing the pump power. The larger the laser gain, the more solitons would be formed in the cavity.

When the linear cavity phase delay bias is set at a very large value, say at about $\delta\Phi_l=1.8\pi$, which is still in the positive cavity feedback range but close to the other end, no stable propagation of the solitons in cavity can be obtained. With the linear cavity phase delay bias selection, there is a big difference between the linear cavity loss and the nonlinear cavity loss. Therefore, if the gain of the laser is smaller than the dynamical loss that a soliton experienced, the soliton quickly dies out as shown in fig. 7a. While if the gain of the laser is even slightly larger than the dynamical loss that a soliton experienced, the soliton peak power will increase. Higher soliton peak power results in smaller dynamical loss and even larger effective gain, therefore, the soliton peak will continuously increase. Associated with the soliton peak increase the soliton pulse width decreases, eventually the soliton breaks up into two solitons with weak peak power and broad pulse width as described by Kärtner *et. al.* [14]. Once a soliton is broken into two solitons with weak peak power, the dynamical loss experienced by each of the solitons becomes very big. Consequently the gain of the laser cannot support them. The new solitons are then immediately destroyed as shown in fig. 7b. If very large gain is available in the laser, the new solitons may survive in the cavity temporally and each of them repeats the same process as shown in fig. 7b, and eventually a state as shown in fig. 7c is formed. Therefore, no stable soliton propagation is possible with too large linear cavity phase delay setting in the laser.

Even in the cases of stable multiple soliton operation, depending on the selection of the linear cavity phase delay, the solitons obtained have different parameters. Fig. 8 shows for comparison the multiple soliton operation obtained with the linear cavity phase delay bias set at $\delta\Phi_l=1.55\pi$. It is to see that solitons with higher peak power and narrower pulse width can be formed with the linear cavity phase delay setting. Extensive numerical simulations have shown that the larger the linear cavity phase delay setting, the higher the soliton peak and the narrower of the soliton pulse achievable.

## 5. Mechanism of the multiple soliton generation and soliton energy quantization

Apparently, depending on the laser linear cavity phase delay bias setting, there exist two different mechanisms of new soliton generation in the laser. One is the soliton shaping of the unstable dispersive waves or the CW components, and the other one is the well-known mechanism of pulse splitting. It is to see that in the laser the process of soliton splitting occurs only in the regime where the new solitons formed are practically unstable. We have already reported previously the phenomenon of soliton generation through unstable background in the lasers [33]. Here we further explain its physical origin.

Our soliton fiber laser is mode locked by using the NPR technique. The operation mechanism of the technique has already been analyzed by several authors [24,34,35]. It has been shown that through inserting a polarizer in the cavity and appropriately setting the linear cavity phase delay, the NPR could generate an artificial saturable absorption effect in the laser. It is the artificial saturable absorber effect that causes the self-started mode locking of the laser. And after a soliton is formed in the laser cavity, it further stabilizes the soliton. Although previous studies have correctly identified the effects of NPR and the saturable absorber in the laser, there is no further analysis on how and to what extent these effects affect the soliton parameters and soliton dynamics. Here we follow the description of C. J. Chen et al [34] to complete it. Our approach is to first determine the linear and nonlinear cavity transmission of the laser, and then based on the results to further find out how they affect the solitons formed in the laser. Physically, the laser cavity shown in Fig. 1 can be simplified to a setup as shown in figure 9 for the purpose of determining its transmission property. Starting from the intracavity polarizer, which sets the initial polarization of light in relation to the birefringent axes of the fiber, the polarization of light after passing through the fiber is determined by both the linear and nonlinear birefringence of the fiber. The light finally passes through the analyzer, which in the experimental system is the same intracavity polarizer. If we assume that the polarizer has an orientation of θ angle with respect to the fast axis of the fiber, and the analyzer has an angle of φ, the phase delay between the two orthogonal polarization components caused by the linear fiber birefringence is $\Delta\Phi_l$, and caused by the nonlinear birefringence is $\Delta\Phi_{nl}$, it can be shown that the transmission coefficient of the setup or the laser cavity is [34]

$$T = \sin^2(\theta)\sin^2(\varphi) + \cos^2(\theta)\cos^2(\varphi) + \frac{1}{2}\sin(2\theta)\sin(2\varphi)\cos(\Delta\Phi_l + \Delta\Phi_{nl}) \qquad (3)$$

C. J. Chen *et. al.* [34] and R.P. Davey *et. al.* [35] have already shown how to select the orientations of the polarizer and the analyzer so that the cavity would generate efficiently saturable absorption effect. In a previous paper [25] we have also shown that the linear cavity transmission of the laser is a sinusoidal function of the linear cavity phase delay $\Delta\Phi_l$ with a period of 2π. It is to point out that within one period of the linear cavity phase delay change, the laser cavity can provide positive (the saturable absorber type) cavity feedback only in half of the period, in the other half of the period it actually has negative feedback.

As shown in equation (3), the actual cavity transmission for an optical pulse is also the nonlinear phase delay $\Delta\Phi_{nl}$ dependent. To illustrate functions of this part we use our simulations as an example. In our simulations the orientation of the polarizer has an angle of θ = 0.125π to the fast axis of the fiber, so light propagation in the fiber will generate a negative nonlinear phase delay. The linear cavity beat length is ¼ of the cavity length, therefore, the maximum linear cavity transmission is at the positions of $(2n+1)\pi$ linear cavity phase delays, where n=0,1,2…. Furthermore, when the linear cavity phase delay is biased within the range between the $(2n+1)\pi$ to $2(n+1)\pi$, the cavity will generate a positive feedback, as under the effect of the nonlinear polarization rotation the actual cavity transmission increases. While if the linear cavity phase delay is located in the range from $2n\pi$ to $(2n+1)\pi$, the cavity will generate a negative feedback. The maximum linear cavity transmission point also marks the switching position of the two feedbacks. For the soliton operation the laser is always initially biased in the positive cavity feedback regime. It is clearly to see that depending on the selection of the linear cavity phase delay and the strength of the nonlinear phase delay, the cavity feedback is possible to be dynamically switched from the positive feedback regime to the negative feedback regime. For the soliton operation of a laser this cavity feedback switching has the consequence that the peak of a soliton formed in the cavity is limited. We found that it is this soliton peak limiting effect that results in the multiple soliton generation in the soliton fiber laser and the soliton energy quantization.

To explain these, we assume that the peak power of a soliton is so strong that it switches the cavity from the positive to the negative feedback regime. In this case although increasing the pump strength will still cause the peak power of the soliton to increase, the higher the soliton peak power increases, the smaller the actual cavity transmission becomes. To a certain fixed value of the soliton peak power, which depends on the linear cavity phase delay setting, further increase of the soliton peak power would results in that the actual cavity transmission that the soliton experiences becomes smaller than the linear cavity transmission. At this point the soliton peak will be clamped. Further increasing the laser gain will not amplify the soliton but the background noise such as the dispersive waves. If the background noise of a certain frequency fulfills the lasing condition, it could also start to lase and form a CW component in the soliton spectrum. We note that coexistence of solitons with CW is a generic effect of the soliton fiber lasers, and the phenomenon was reported by several authors [36, 37]. Linear waves are intrinsically unstable in the cavity due to the modulation instability. When they are strong enough, they become modulated. And under the effect of saturable absorption, the strongest background pulse will be amplified and shaped into a new soliton. This was exactly what we have observed in the experiments on how a new soliton was generated. The two solitons in the cavity share the same laser gain. As the cavity generates a positive feedback for the weak soliton and a negative one for the strong soliton, under the gain competition the two solitons have to adjust their strength so that the stronger one becomes weaker, and the weaker one becomes stronger, eventually they will stabilize at a state that both solitons have exactly the same peak power. The soliton internal energy balance further determines their other parameters. Except that there are interactions between the solitons, they will always have identical parameters in the stable state.

It turns out that the multiple soliton formation in the laser is in fact caused by the peak power clamping effect of the cavity. In addition, the soliton energy quantization observed is also a nature consequence of the gain competition between the solitons in the laser. Obviously the maximum achievable soliton peak power in the laser is the linear cavity phase delay dependent. When the linear cavity phase delay is set close to the cavity feedback switching point, solitons with relatively lower peak power could already dynamically switch the cavity feedback. Therefore, solitons obtained at this linear cavity phase delay setting have lower peak power and broader pulse width as shown in fig. 6a. While if the linear cavity phase delay is set far away from the switching point, soliton peak power is clamped at a higher value, solitons with higher peak and narrower pulse width would then be obtained as shown in fig. 8. In particular, if the linear cavity phase delay is set too close to the switching point, as the peak power of the pulse is clamped to too small value, except the mode-locked pulses, no soliton could be formed in the laser. While when the linear cavity phase delay is set too far away from the switching point as demonstrated numerically in figure 7, before the soliton peak reaches the switching point, it has already become so high and so narrow that it splits, no stable soliton propagation could be obtained in the laser. Instead only the state of so-called noise-like pulse emission will be observed [38].

Finally we note that the multiple soliton operation and soliton energy quantization effect have also been observed in other passively mode-locked soliton fiber lasers, such as in the figure-of-eight lasers and the lasers passively mode locked with semiconductor saturable absorbers [8,9]. Even in the actively mode-locked fiber lasers [2] these phenomena have also been observed. Despite of the fact that those soliton lasers are not mode-locked with the NPR technique, therefore, their detailed cavity transmission could not have the same feature as described by equation (3), enlightened by the result obtained in our laser, we conjugate that there must also have a certain pulse peak power limiting mechanism in the lasers, which causes their multiple pulse formation. Indeed, we found that for the figure-of-eight lasers, if the fiber birefringence of the nonlinear loop is further considered, it would also generate a similar pulse peak clamping effect in the laser. However, birefringence of fibers in the lasers is normally ignored. It was also reported that due to the two-photon absorption effect the SESAM used for the passive mode-locking of fiber lasers has a pulse peak power limiting effect [39]. It is therefore not surprising that soliton laser mode-locked with the material could also exhibit multiple solitons. For the actively mode-locked laser, in most cases the multiple soliton generation is due to the harmonic mode locking. In this case as too many solitons share the limited cavity gain, the energy of each pulse is weak. Therefore, even when the net cavity dispersion is negative, solitons are normally difficult to be form. We point out that for an actively mode-locked fiber laser if the cavity is not carefully designed, the cavity birefringence combined with the modulator, which is a polarizing device, could form a birefringence filter and further limit the peak power of the pulses formed in the lasers.

## 5. Conclusions

In conclusion, we have numerically studied the mechanism of multiple soliton generation and soliton energy quantization in a soliton fiber ring laser passively mode locked by using the nonlinear polarization rotation technique. We identified that the multiple soliton generation in the laser is caused by the peak power clamping effect of the cavity. Depending on the linear cavity phase delay setting, the nonlinear phase delay generated by a soliton propagating in the fiber cavity could be so large that it switches the cavity feedback from the initially selected positive regime into the negative regime. And as a result of the cavity feedback change the maximum achievable soliton peak power is then limited. In this case increasing the laser pump power will not increase the peak power of the solitons, but generate a new soliton. Therefore, multiple solitons are formed in the laser. As the solitons share the same laser gain, gain competition between them combined with the cavity feedback feature further results in that in the steady state they have exactly the same soliton parameters. The parameters of solitons formed in the laser are not fixed by the laser configuration but vary with the laser operation conditions, which are determined by the soliton internal energy balance between the shared laser gain and the dynamical losses of each soliton.

**References:**

1. K. Smith, J. R. Armitage, R. Wyatt, N. J. Doran, "Erbium fiber soliton laser", Electron. Lett., 26, 1149-1151(1990).

2. R. P. Davey, N. Langford, A. I. Ferguson, "Interacting solitons in erbium fiber laser", Electron. Lett., 27, 1257-1259(1991).

3. D. J. Richardson, R. I. Laming, D. N. Payne, V. J. Matsas, M. W. Phillips, "Pulse repetition rates in passive, femtosecond soliton fiber laser", Electron. Lett., 27, 1451-1453( 1991).

4. M. Nakazawa, E. Yoshida and Y. Kimura, "Generation of 98fs optical pulses directly from an erbium doped fiber ring laser at 1.57 um", Electron. Lett., 29, 63-65(1993).

5. V. J. Matsas, D. J. Richardson, T. P. Newson and D. N. Payne, "Characterization of a self-starting, passively mode-locked fiber ring laser that exploits nonlinear polarization evolution", Opt. Lett., 18, 358-360 (1993).

6. K. Tamura, E. P. Ippen, H. A. Haus and L. E. Nelson, "77-fs pulse generation from a stretched-pulse mode-locked all-fiber ring laser", Opt. Lett., 18, 1080-1082(1993).

7. M. Nakazawa, E. Yoshida, T. Sugawa and Y. Kimura, "Continuum suppressed, uniformly repetitive 136fs pulse generation from an erbium doped fiber laser with nonlinear polarization rotation", Electron. Lett., 29, 1327-1329(1993).

8. L. A. Gomes, L. Orsila, T. Jouhti and O. G. Okhotnikov, "Picosecond SESAM-based ytterbium mode-locked fiber lasers", IEEE J. of selected topics in quantum Electronics, 10, 129-136 (2004).

9. B. C. Collings, K. Bergman, S. T. Cundiff, S. Tsuda, J. N. Kutz, J. E. Cunningham, W. Y. Jan, M. Koch and W. H. Knox, "Short cavity erbium/ytterbium fiber lasers mode-locked with a saturable Bragg reflector", IEEE Journal of Selected Topics in Quantum Electronics, 3, 1065-1075(1997).

10. A. B. Grudinin, D. J. Richardson and D. N. Payne, "Energy quantisation in figure eight fibre laser", Electron. Lett. 28, 67-68 (1992).

11. N. N. Akhmediev, A. Ankiewicz, J. M. Soto-Crespo, "Multisoliton solutions of the complex Ginzburg-Landau equation", Phys. Rev. Lett., 79, 4047-4051 (1997).

12. M. J. Lederer, B. Luther-Davies, H. H. Tan, C. Jagadish, N. N. Akhmediev and J. M. Soto-Crespo, "Multipulse operation of a Ti:sapphire laser mode locked by an ion-implanted semiconductor saturable-absorber mirror", J. Opt. Soc. Am. B, 16, 895-904, 1999.

13. C. Spielmann, P. F. Curley, T. Brabec, and F. Krausz, "Ultrabroadband femtosecond lasers", IEEE J. Quantum Electron., 30, 1100-1114(1994).

14. F. X. Kärtner, J. Aus der Au, and U. Keller, "Mode-locking with slow and fast saturable absorbers – what's the difference?", IEEE J. of selected Topics in Quantum Electronics, 4, 159-168(1998).

15. G. P. Agrawal, "Optical pulse propagation in doped fiber amplifiers", Phys. Rev. A., 44, 7493-7501(1991).

16. P. Grelu, F. Belhache, F. Gutty and J. M. Soto-Crespo, "Relative phase locking of pulses in a passively mode-locked fiber laser", J. Opt. Soc. Am. B, 20, 863-870(2003).

17. Ph Grelu and J. M. Soto-Crespo, "Mutlisoliton states and pulse fragmentation in a passively mode-locked fiber laser", J. Opt. B: Quantum Semiclass. Opt.,6, S271-S278(2004).

18. A. Komarov, H. Leblond and F. Sanchez, "Multistability and hysteresis phenomena in passively mode-locked fiber lasers", Phy. Rev. A, 71, 053809(2005).

19. D. Y. Tang, W. S. Man and H. Y. Tam, "Stimulated soliton pulse formation and its mechanism in a passively mode-locked fiber soliton laser", Opt. Commun., 165, 189-194(1999).

20. B. Zhao, D. Y. Tang, P. Shum, Y. D. Gong, C. Lu, W. S. Man and H. Y. Tam, "Energy quantization of twin-pulse solitons in a passively mode-locked fiber ring laser", Appl. Phys. B., 77, 585-588(2003).

21. D. Y. Tang, W. S. Man, H. Y. Tam and P. Drummond, "Observation of bound states of solitons in a passively mode-locked fiber soliton laser", Phys. Rev. A, 66, 033806 (2002).

22. M. Nakazawa, E. Yoshida and Y. Kimura, "Low threshold 290 fs erbium-doped fiber laser with a nonlinear amplifying loop mirror pumped by InGaAsP laser diodes", Appl. Phys. Lett., 59, 2073-2075 (1991).

23. A. K. Komarov and K. P. Komarov, "Multistability and hysteresis phonmena in passive mode-locked lasers", Phys. Rev. E, 62, R7607-R7610(2000).

24. H. A. Haus, J. G. Fujimoto, and E. P. Ippen, "Structures for additive pulse mode-locking", J. Opt. Soc. B, 8, 2068-2076(1991).

25. W. S. Man, H. Y. Tam, M. S. Demonkan, P. K. A. Wai and D. Y. Tang, "Mechanism of intrinsic wavelength tuning and sideband asymmetry in a passively mode-locked soliton fiber ring laser", J. Opt. Soc. B, 17, 28-33(2000).

26. D. Y. Tang, S. Fleming, W. S. Man, H. Y. Tam and M. S. Demokan, "Subsideband generation and modulational instability lasing in a fiber soliton laser", J. Opt. Soc. Am. B, 18, 1443-1450(2001).

27. D. Y. Tang, B. Zhao, D. Y. Shen, C. Lu, W. S. Man and H. Y. Tam, "Compound pulse solitons in a fiber ring laser", Phys. Rev. A, 68, 013816(2003).

28. B. Zhao, D. Y. Tang, L. M. Zhao, P. Shum and H. Y. Tam, "Pulse train nonuniformity in a fiber soliton ring laser mode-locked by using the nonlinear polarization rotation technique", Phys. Rev. A, 69, 43808(2004).

29. G. P. Agrawal, "Nonlinear fiber optics", pp51-55, 3$^{rd}$ edition, Academic Press, New York, 2001.

30. V. S. Grigoryan and T. S. Muradyan, "Evolution of light pulses into autosolitons in nonlinear amplifying media", J. Opt. Soc. Am. B, 8, 1757-1765(1991).

31. S. M. J. Kelly, "Characteristic sideband instability of periodically amplified average soliton", Electron. Lett., 28, 806-807(1992).

32. M. L. Dennis and I. N. Duling III, "Experimental study of sideband generation in femtosecond fiber lasers", IEEE J. Quantum Electron., 30, 1469-1477(1994).

33. W. S. Man, H. Y. Tam, M. S. Demonkan and D. Y. Tang, "Soltion shaping of dispersive waves in a passively node-locked fiber soliton ring laser", Optical and Quantum Electronics, 33, 1139-1147(2001).

34. C. J. Chen, P. K. A. Wai and C. R. Menyuk, "Soliton fiber ring laser", Opt. Lett., 17, 417-419(1992).

35. R. P. Davey, N. Langford and A. I. Ferguson, "Role of polarization rotation in the modelocking of an Er fiber laser", Electron. Lett., 29, 758-760(1993).

36. S. Namiki, E. P. Ippen, H. A. Haus and K. Tamura, "Relaxation oscillation behavior in polarization additive pulse mode-locked fiber ring lasers", Appl. Phys. Lett., 69, 3969-3971(1996).

37. B. Zhao, D. Y. Tang and H. Y. Tam, "Experimental observation of FPU recurrence in a fiber ring laser", Opt. Express, 11, 3304-3309(2003).

38. M. Horowitz, Y. Barad and Y. Silberberg, "Noiselike pulses with broadband spectrum generated from an erbium-doped fiber laser", Opt. Lett., 22, 799-801 (1997).

39. E. R. Thoen, E. M. Koontz, M. Joschko, P. Langlois, T. R. Schibli, F. X. Kartner, E. P. Ippen and L. A. Kolodziejski, "Two-photon absorption in semiconductor saturable absorber mirrors", Appl. Phys. Lett., 74, 3927-3929(1999).

**Figure captions:**

Fig. 1: A schematic of the soliton fiber laser. PI: Polarization dependent isolator. PC: Polarization controller. DSF: Dispersion shifted fiber. EDF: Erbium-doped fiber. WDM: Wavelength-division-multiplexer.

Fig. 2: A typical experimentally measured oscilloscope trace of the multiple soliton operation of the laser.

Fig. 3: Numerically calculated multiple soliton operation state of the laser. $\delta\Phi_l = 1.20\pi$, G=350. Other parameters used are described in the text.

Fig. 4: Relationship between the soliton number in the simulation window and the pump strength. $\delta\Phi_l = 1.20\pi$.

Fig. 5: Soliton shaping of the mode-locked pulse in the laser. $\delta\Phi_l = 1.20\pi$. Top figure: Evolution of pulse profile with the pump strength. Bottom figure: Evolution of the optical spectra with the pump strength.

Fig. 6: Process of the new soliton generation in the laser. $\delta\Phi_l = 1.20\pi$. a) G=255. b) G=270. c) G=275.

Fig. 7: Soliton evolutions calculated with $\delta\Phi_l = 1.80\pi$. a): G=470. b) G=478. c) G=600.

Fig. 8: Multiple soliton operation of the laser calculated with linear cavity phase delay bias set at $\delta\Phi_l = 1.55\pi$, G = 465.

Fig. 9: An equivalent setup of Fig. 1 for determining the cavity transmission.

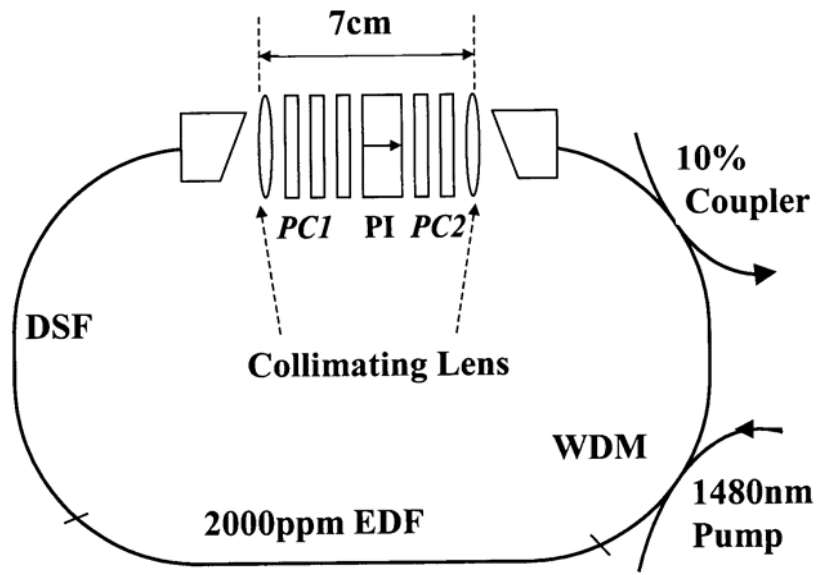

Fig. 1.

D. Y. Tang et. al.

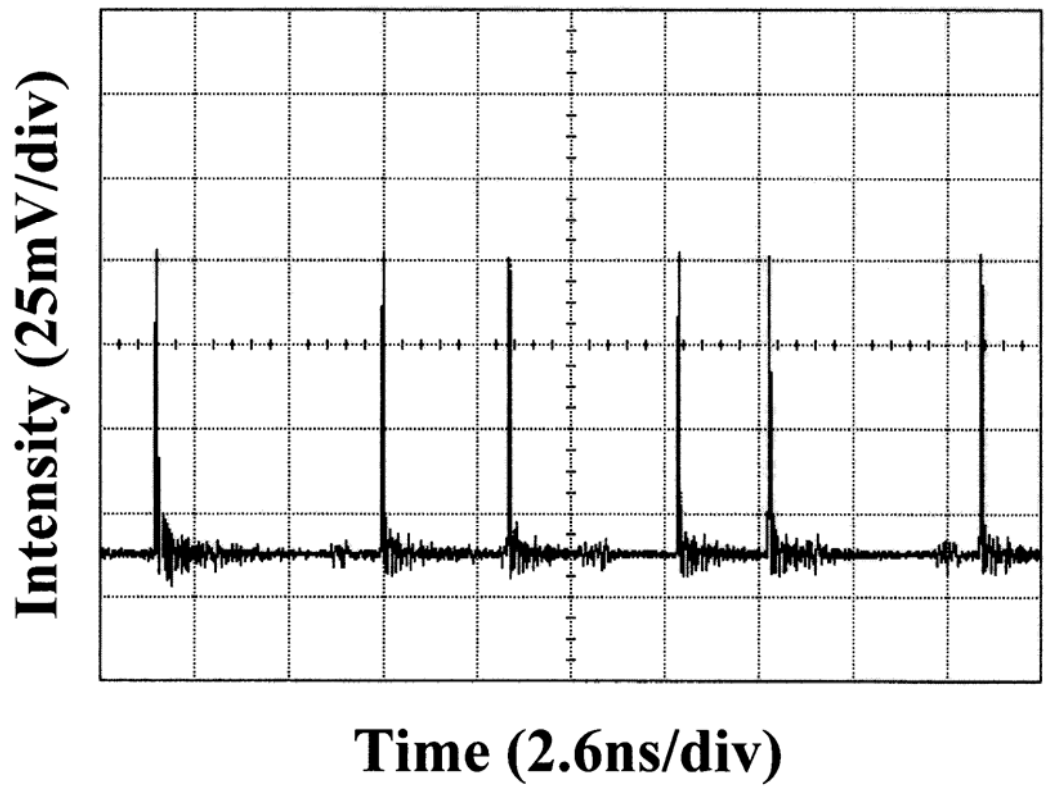

Fig. 2

D. Y. Tang et. al.

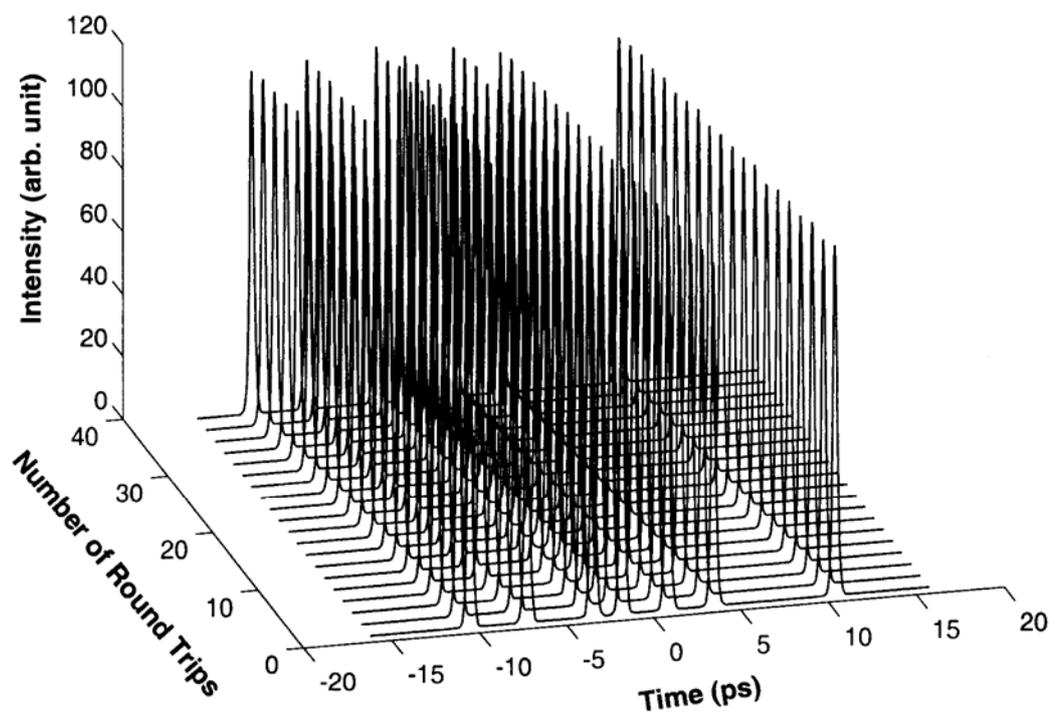

Fig. 3.

D. Y. Tang et. al.

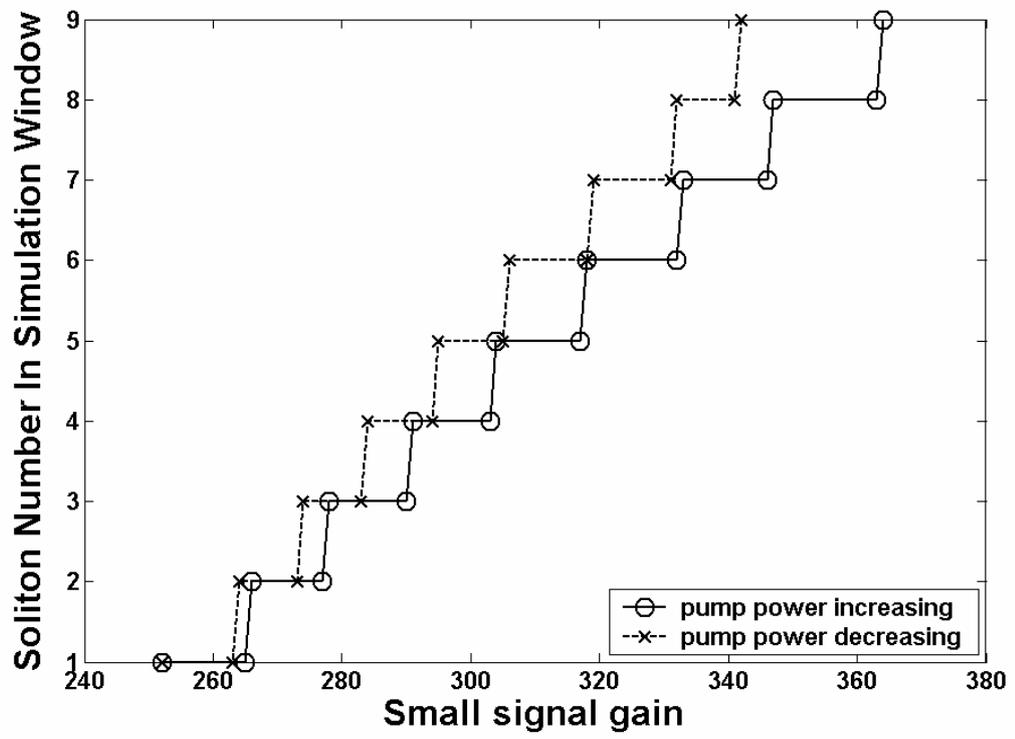

Fig. 4.

D. Y. Tang et. al.

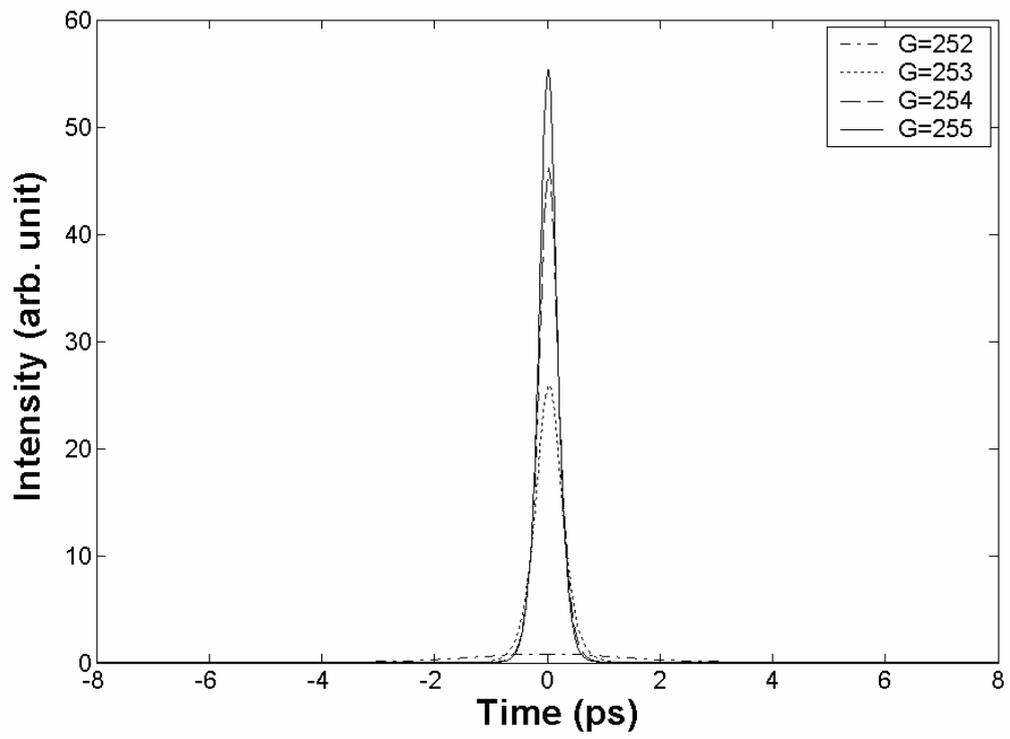

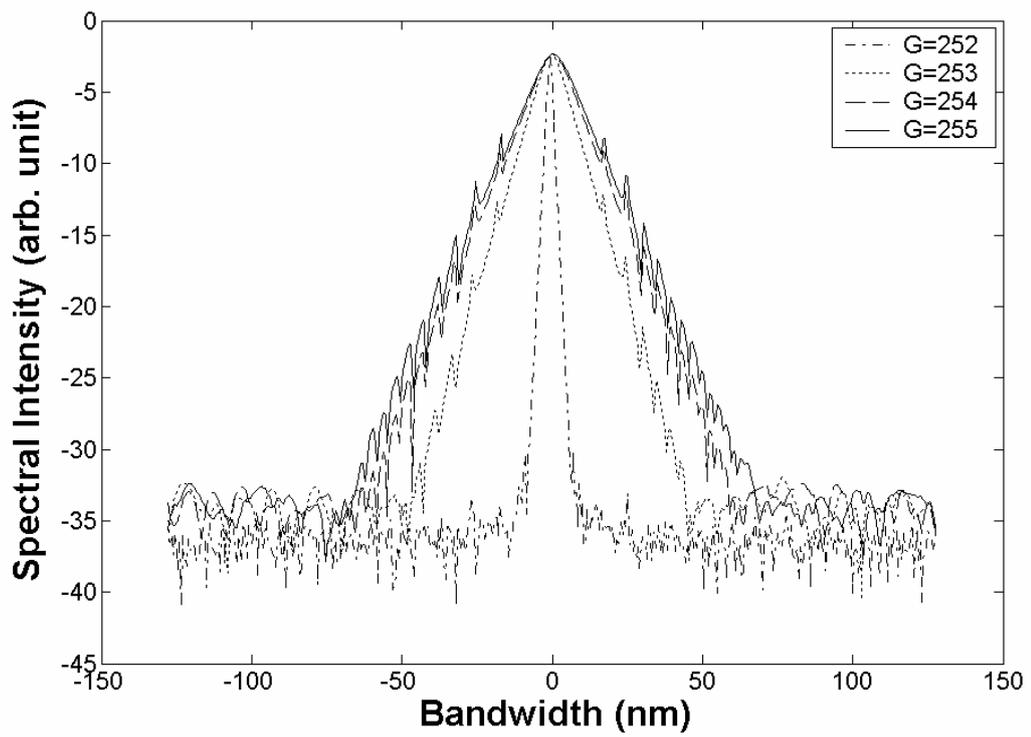

Fig. 5.

D. Y. Tang et. al.

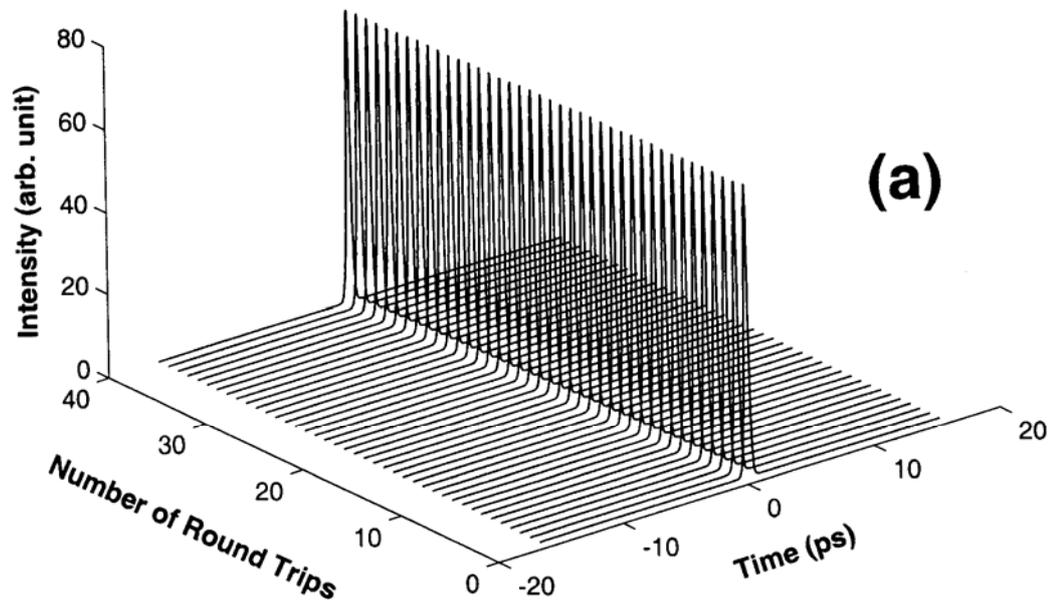

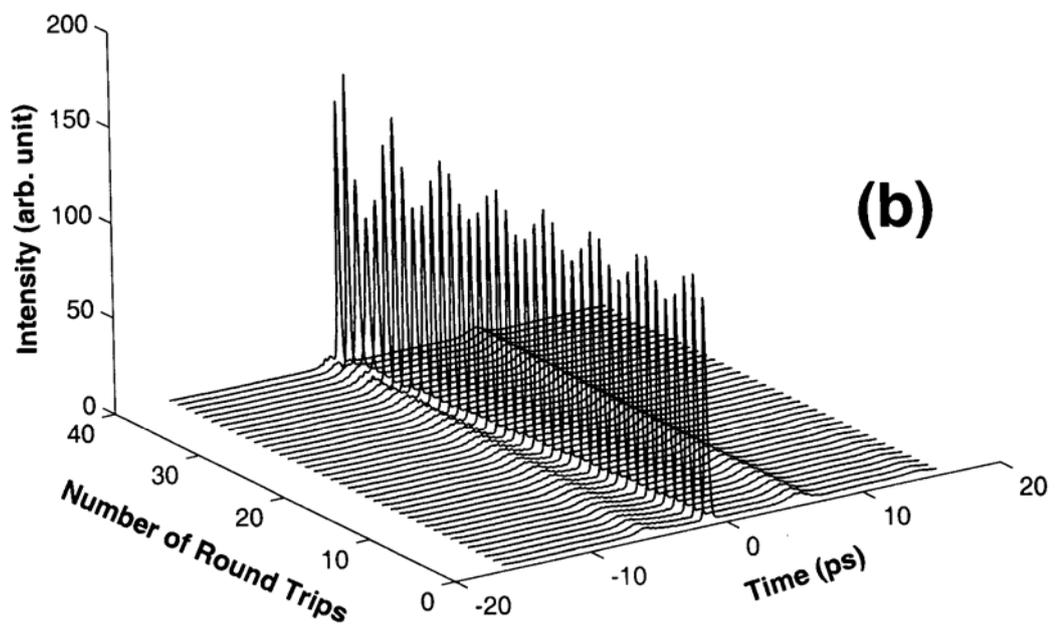

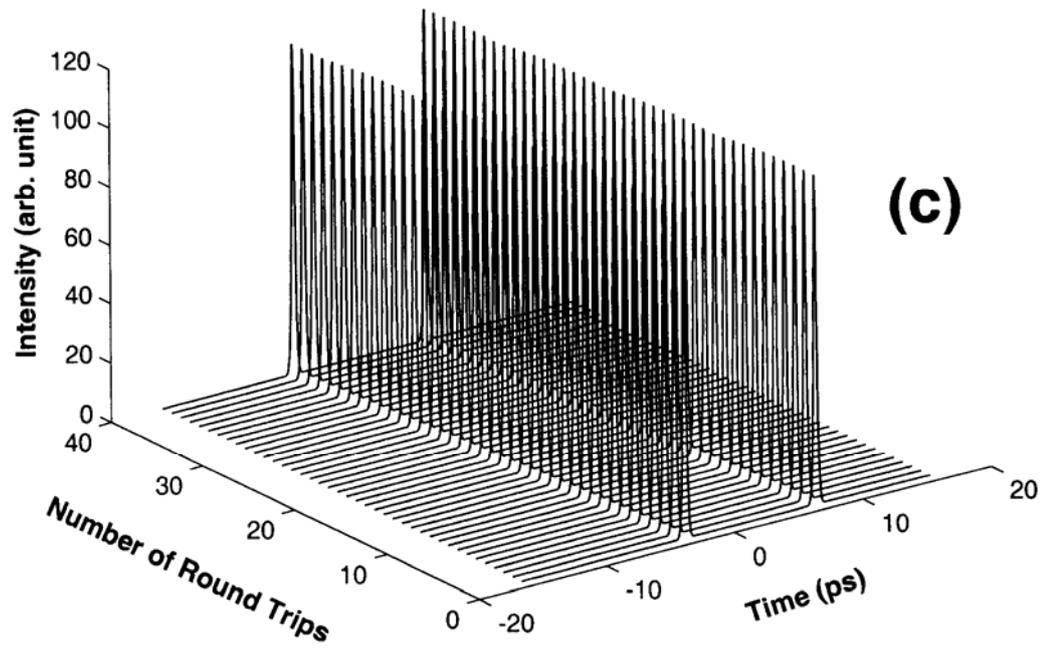

Fig.6

D. Y. Tang et. al.

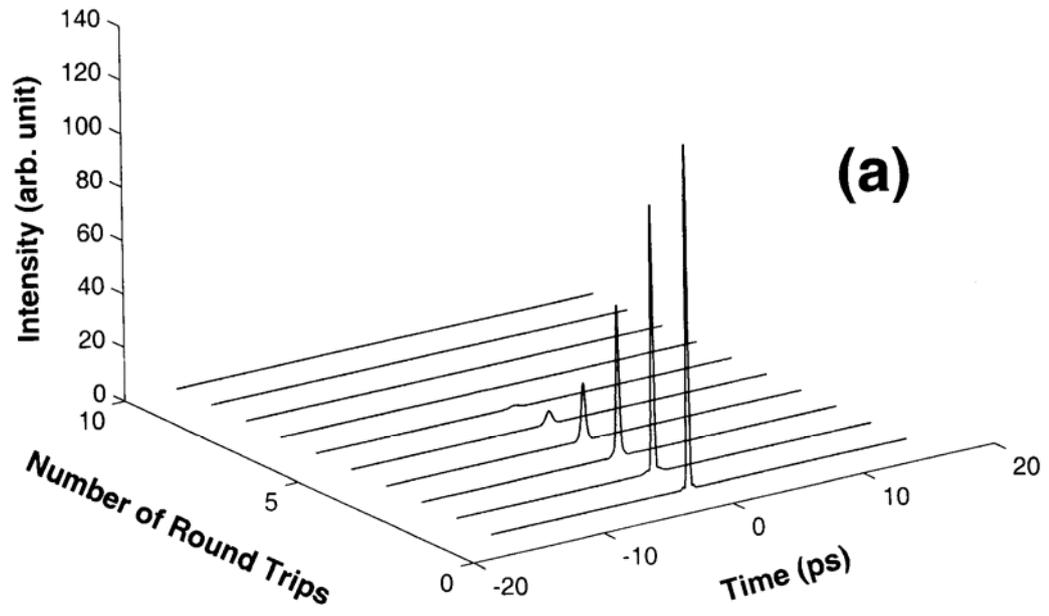

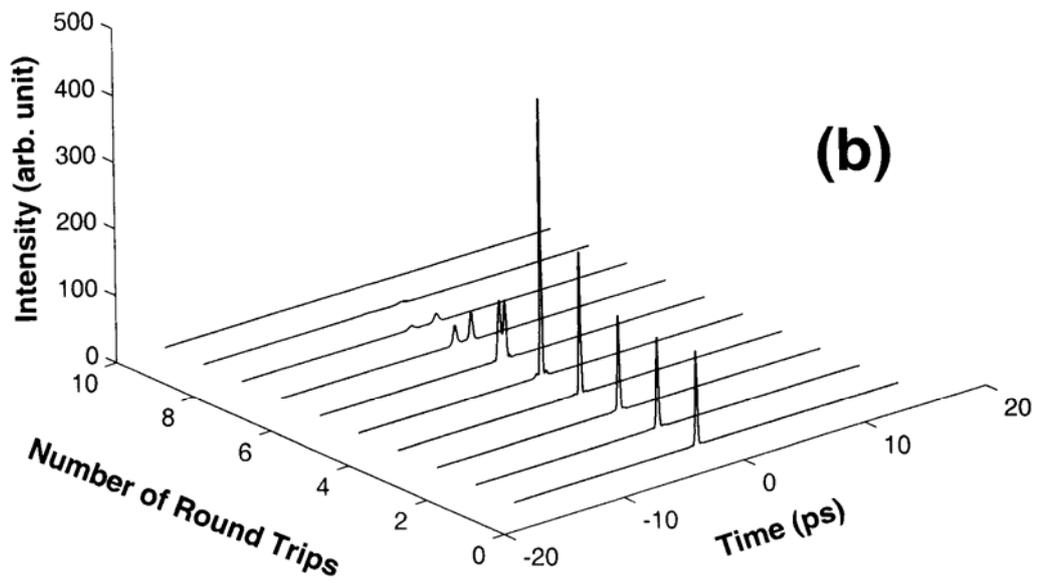

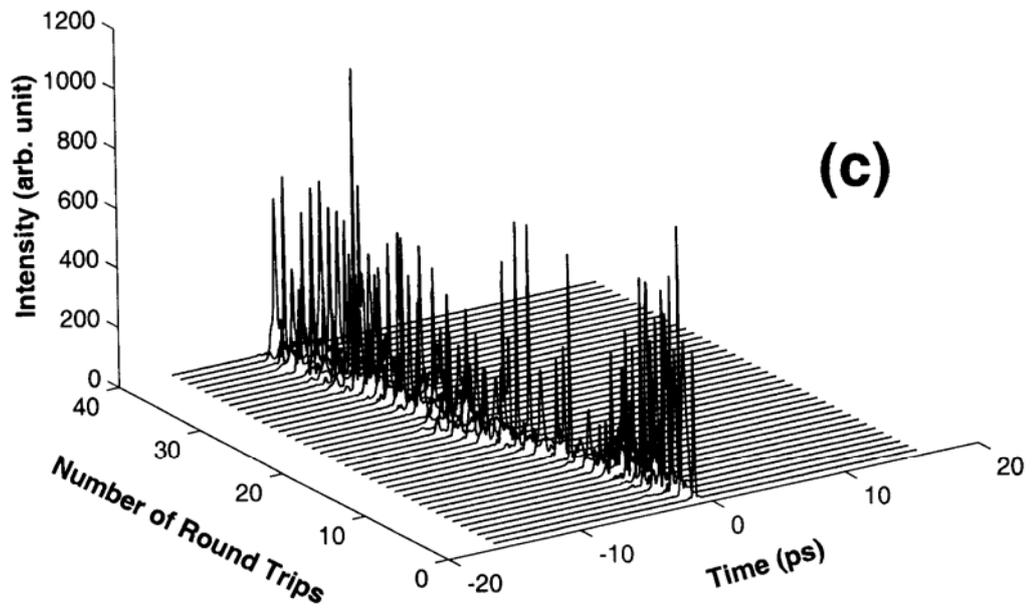

Fig. 7

D. Y. Tang et. al.

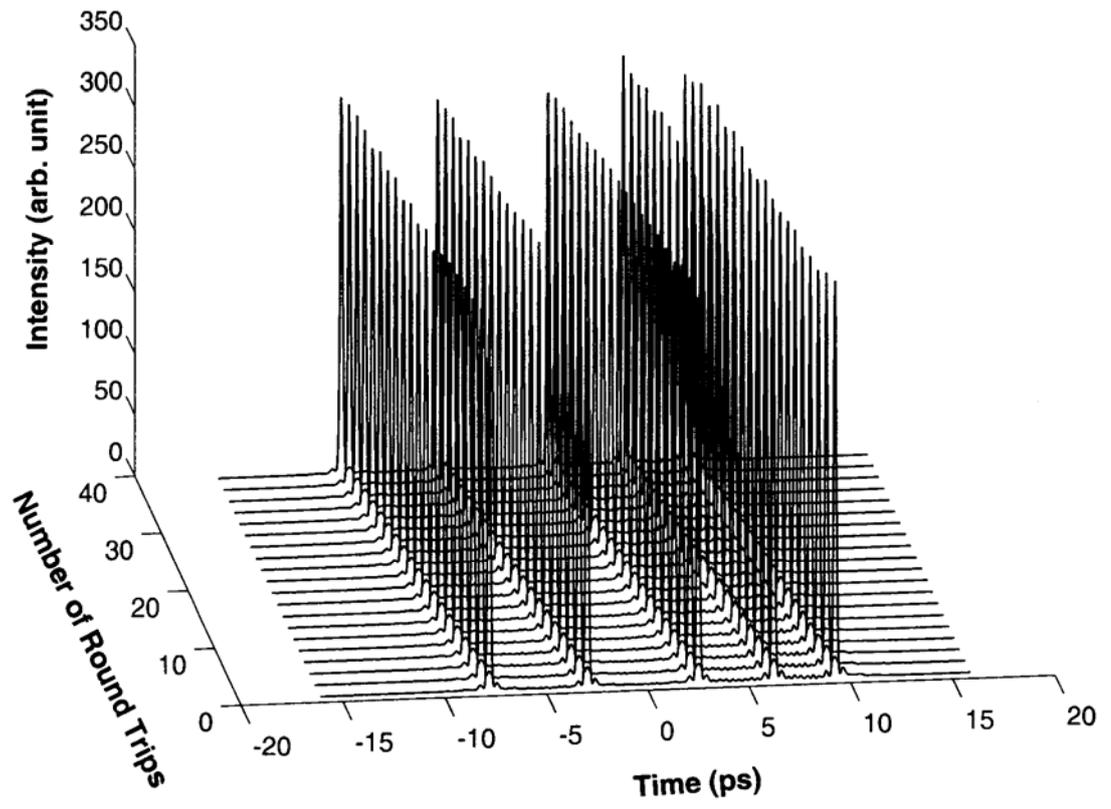

Fig. 8

D. Y. Tang et. al.

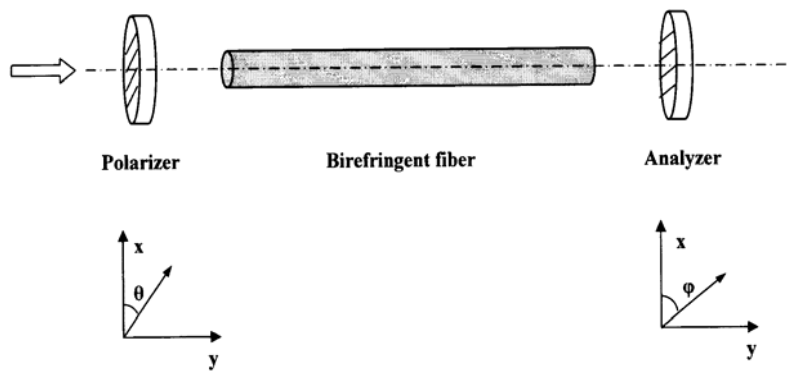

Fig. 9.

D. Y. Tang et. al.